%% file: main.tex
\documentclass[preprint, 12pt]{elsarticle}
\pdfoutput=1

\usepackage[american]{babel}
\usepackage[T1]{fontenc}
\usepackage[utf8]{inputenc}
\usepackage{amsmath}
\usepackage{amssymb}
\usepackage{amsthm}
\usepackage{mathtools}
\usepackage{bbm}
\usepackage{mleftright}
\usepackage{xfrac}
\usepackage{tensor}
\usepackage{fnbreak}
\usepackage{listingsutf8}
\usepackage{lstautogobble}
\usepackage{textgreek}
\usepackage{booktabs}
\usepackage{caption}
\usepackage{graphicx}
\usepackage{gincltex}
\usepackage{xcolor}
\usepackage{tikz}
\usepackage[useregional, en-GB]{datetime2}
\usepackage{csquotes}
\usepackage{microtype}
\usepackage[unicode]{hyperref}
\usepackage{cleveref}

\usetikzlibrary{calc, positioning}
\AtBeginDocument{\DTMsetstyle{en-GB}}

\input{settings.tex}
\input{listings_styles.tex}

\journal{Computer Physics Communications}

\begin{document}

\begin{frontmatter}

\title{\minlag: Generating and studying dark matter model Lagrangians with just the particle content}

\author[a]{Simon May\corref{author}}
\ead{simon.may@mpa-garching.mpg.de}
\ead[url]{https://www.mpa-garching.mpg.de/~smay/}

\cortext[author]{Corresponding author.}
\address[a]{Max-Planck-Institut für Astrophysik, Karl-Schwarzschild-Straße 1, 85741 Garching, Germany}

\begin{abstract}
	{\minlag} is a {\Python} program which allows one to specify the field content of an extension of the Standard Model of particle physics and, using this information, to generate the most general renormalizable Lagrangian that describes such a model.
	As the program was originally created for the study of minimal dark matter models with radiative neutrino masses, it can handle additional scalar or Weyl fermion fields which are $\SUIIIC$ singlets, $\SUIIL$ singlets, doublets or triplets, and can have arbitrary $\UIY$ hypercharge.
	It is also possible to enforce an arbitrary number of global $\gU(1)$ symmetries (with $\gZ_2$ as a special case) so that the new fields can additionally carry such global charges.
	In addition to human-readable and {\LaTeX} output, the program can generate {\SARAH} model files containing the computed Lagrangian, as well as information about the fields after electroweak symmetry breaking (EWSB), such as vacuum expectation values (VEVs) and mixing matrices.
	This capability allows further detailed investigation of the model in question, with {\minlag} as the first component in a tool chain for rapid phenomenological studies of \q{minimal} dark matter models requiring little effort and no unnecessary input from the user.
\end{abstract}


\end{frontmatter}

\noindent
{\bf PROGRAM SUMMARY}

{\small
\noindent
{\em Program Title:} \minlag
\\
{\em Licensing provisions:} GPLv3
\\
{\em Programming language:} \Python
\\

\noindent
{\em Nature of problem:}\\
	Given a quantum field theory's gauge group, it is sufficient to specify the particle (field) content in order to identify the full renormalizable theory, up to the parameters in its Lagrangian.
	However, the process of determining the Lagrangian manually is not only tedious and error-prone, but also involves additional complications such as redundant terms or the question of whether the theory is anomaly-free.
\\
{\em Solution method:}\\
	{\minlag} generates the complete renormalizable Lagrangian for a given model with the Standard Model gauge group $\SUIIIC \times \SUIIL \times \UIY$, including interaction terms.
	Redundant terms in the Lagrangian are eliminated in order to avoid duplicated parameters.
	The particle content is also checked for gauge anomalies, including the Witten $\gSU(2)$ anomaly \cite{fm:witten-su2}.
	The model will automatically be modified to make fermions vector-like if necessary.
	The generated Lagrangian can be output in {\SARAH} \cite{fm:sarah4,fm:sarah-summary} model file format so that the model is immediately available for detailed phenomenological study using the capabilities of {\SARAH}.
\\
{\em Additional comments including restrictions and unusual features:}\\
	Instead of manually determining the details of a model, the only input to the program {\minlag} is the particle content.
	Using the output to {\SARAH}, {\minlag} thus forms the first step in a tool chain which enables the complete implementation and study of a new model with minimal effort and no \q{boilerplate} user input.
	The focus is on \q{minimal} dark matter models, i.\,e.\ those with the Standard Model gauge group (no additional gauge fields), where the new fields are color singlets and at most triplets under $\SUIIL$.
\\

}

\section{Introduction}
The program {\minlag} is able to generate the most general renormalizable Lagrangian describing certain classes of extensions of the Standard Model (SM) of particle physics in a fully automatic fashion, with several different output formats, requiring only the model's field content to do so.
As the program was originally written for the study of minimal dark matter models with radiative neutrino masses \cite{may2020minimal}, it can handle fields beyond the Standard Model (BSM) with the following properties:
\begin{itemize}
	\item scalar or fermion fields\footnote{%
		Fermions are always defined in terms of Weyl spinors, cf.\ \ref{app:conventions}.%
	}
	\item $\gSU(3)$ singlets
	\item $\gSU(2)$ singlets, doublets or triplets
	\item arbitrary $\gU(1)$ hypercharge
	\item charged under an arbitrary number of global $\gU(1)$ symmetries (with $\gZ_2$ as a special case)
\end{itemize}
{\minlag} originated as an effort to study the minimal dark matter models with radiative neutrino masses introduced in \cite{Restrepo:2013aga} in a general way.
A missing piece in \cite{Restrepo:2013aga} are the Lagrangians for the individual models, which are only specified via the field content for each model.

While it is a manageable task for someone experienced in building particle physics models to construct the Lagrangian for such a model manually, this process can be error-prone and time-consuming – not only because all possible terms must be enumerated without omissions, but especially when taking into account gauge anomalies and various identities for the possible terms (cf.\ \cref{app:identities}).
Moreover, since the symmetry groups and the kinds of representations used in these models are limited and fixed, it is quite feasible to exhaustively list all the terms that could potentially appear in such a Lagrangian in full generality in an automated fashion.
The models are thus checked for gauge anomalies, including the Witten $\gSU(2)$ anomaly \cite{Witten:1982fp}, which are avoided by introducing vector-like fermions if necessary.
Redundant terms are omitted using the identities in \cref{app:identities}.
Even for manually-constructed Lagrangians, {\minlag} can still serve as a useful cross-check in order to verify that there are neither too few (forgotten) nor too many (redundant) terms.

In addition, model files for {\SARAH} \cite{Staub:2013tta,Staub:2015kfa} can be constructed automatically from the specified field content, which can be tedious if done manually.
Thus, {\minlag} enables rapid phenomenological studies using {\SARAH} and, successively, further tools like {\SPheno} \cite{Porod:2003um,Porod:2011nf} and {\micrOMEGAs} \cite{Belanger:2001fz,Belanger:2018mqt}.

For a typical number of fields, which is small in most models ($< 10$), the program's execution time should be of the order of one second or less on any hardware in current use.

\section{Download and installation}
\label{sec:install}
{\minlag} is available on the Python Package Index (PyPI) at \url{https://pypi.org/project/minimal-lagrangians}.
Therefore, the simplest method to obtain and install the program files is to use the {\pip} package manager.
Provided that {\pip} is set up on the machine, running
\begin{lstlisting}[style=display-sh, numbers=none]
	pip install minimal-lagrangians
\end{lstlisting}
on the command line should be all that is necessary to install \minlag.

Further development of the program is tracked using the {\git} version control system.
The most recent development snapshot is available at \url{https://gitlab.com/Socob/minimal-lagrangians}.
Bugs or other issues should be reported there as well.

The program is written using the \Python~3 programming language and should run using {\Python} versions $\ge 3.4$.
It can be run on any system which {\Python} is available for.
Apart from that, there are no external dependencies.

\section{Use of the program}
\subsection{Command-line interface}
Information on how to run the program on the command line can be obtained by running \lstinline[language=sh]|minimal-lagrangians -h|:
\begin{lstlisting}[style=plain-text-80col]
usage: minimal-lagrangians [-h] [--format {LaTeX,SARAH,plain}]
                           [--model-file [path/to/file.py]] [--omit-equivalent-scalars]
                           [--omit-self-interaction] [--list-discarded-terms]
                           [--sarah-no-scalar-cpv] [--version]
                           model [parameter α]

A Python program to generate the Lagrangians for dark matter models

positional arguments:
  model                 name of the model whose Lagrangian is to be generated (specify
                        “list” in order to list all available models)
  parameter α           value of the model parameter α (determines hypercharges of the
                        fields)

optional arguments:
  -h, --help            show this help message and exit
  --format {LaTeX,SARAH,plain}
                        output format for the generated Lagrangian (default: plain)
  --model-file [path/to/file.py]
                        file containing user-defined models; a file is only read if this
                        option is present (default: ./models.py)
  --omit-equivalent-scalars
                        keep only scalar fields from the model which have unique quantum
                        numbers and absolute hypercharge values (omit duplicates)
  --omit-self-interaction
                        omit pure self-interactions of the new fields in the Lagrangian,
                        that is, output only interaction terms which involve both SM and
                        new fields (default: output all terms)
  --list-discarded-terms
                        list redundant terms which were discarded from the Lagrangian
                        due to identities
  --sarah-no-scalar-cpv
                        assume that there is no CP violation causing mixing between
                        scalar and pseudoscalar fields for SARAH output
  --version             show program’s version number and exit
\end{lstlisting}

As shown in the output listing, the general syntax used to generate a model's Lagrangian is \lstinline[language=sh]|minimal-lagrangians <model_name>|.
The names of all available models can be listed using \lstinline[language=sh]|minimal-lagrangians list|.
Models can have an additional hypercharge offset parameter (called $\alpha$ in \cite{Restrepo:2013aga}); if the model in question does not use such a parameter, it should be omitted on the command line.
New models can be defined in a user-supplied file, which can be specified using the option \lstinline[language=sh]|--model-file [path]|.
The details and syntax of defining new models are laid out in \cref{sec:model_definition}.

The output format can be selected using the \lstinline[language=sh]|--format| option; the different formats are detailed in \cref{sec:output}.
In general, {\minlag} will write its output to the standard output stream.
The exception is the {\SARAH} output format, which requires multiple files with specific names and will thus additionally write to a new directory within the current working directory.

As mentioned before, {\minlag} makes use of the identities in \cref{app:identities} to remove redundant terms from the Lagrangian.
The command line flag \lstinline[language=sh]|--list-discarded-terms| causes valid terms which were omitted for this reason to be printed in the output, allowing for cross-checks and comparisons between different sources which may have chosen different parametrizations of the same Lagrangian.

Two additional flags
can be used to obtain a simplified version of the output.
\lstinline[language=sh]|--omit-equivalent-scalars| can be useful for models with a hypercharge parameter (e.\,g.\ \cite{Restrepo:2013aga}; see also \cref{sec:model_definition} for more details on the definition of such models).
Depending on the value of $\alpha$, several scalar multiplets can end up as adjoints of each other (opposite quantum numbers, \q{$\phi_1^\dagger = \phi_2$}) and thus effectively form nothing but multiple generations of the same scalar field.
An example is the model T3-B for $\alpha = -1$ \cite{Restrepo:2013aga}, which is the so-called scotogenic model \cite{Ma:2006km} for a single scalar doublet.
In this case,
the flag reduces the field content to only unique scalar fields, allowing for the study of the most minimal version of the model without having to define each special case separately.
\lstinline[language=sh]|--omit-self-interaction| simply omits all terms which do not involve any SM fields for cases where such self-interactions are not of interest.

\subsection{Model definition}
\label{sec:model_definition}
{\minlag} comes with a number of built-in model definitions, among them all the models listed in \cite{Restrepo:2013aga}.
The program file \path{data.py} contains these definitions and can thus be inspected for examples of the model definition syntax.

Additional new models can be defined in a user-supplied file using the \lstinline[language=sh]|--model-file [path]| command line option.
The default path for the user model file is a file \path{models.py} in the current working directory, which will be used if no path is specified: \lstinline[language=sh]|minimal-lagrangians --model-file|.

The user model file must contain a list of model definitions with valid {\Python} syntax.
Models can be added as entries to the list in the following form:
\begin{lstlisting}[language=Python, style=display-code]
[
  BSMModel('<model_name>', (
  		# list of fields
  		# (type,     name, SU(2) rep., U(1)_Y,  global sym.)
  		# for a scalar, e.g. a Z₂-odd scalar doublet
  		# with hypercharge 1:
  		ScalarField ('S',  2,          Y=1,     z2=-1),
  		# for a fermion, e.g. a Z₂-even fermion singlet
  		# with hypercharge 0:
  		FermionField('F',  1,          Y=0,     z2=1),
  		# …
  	),
  	# optional: parameter values for different hypercharge
  	# assignments (offsets), e.g.
  	(0, 2, …)
  ),
  # …
]
\end{lstlisting}
The program uses the convention where the hypercharge $Y$ is normalized such that the electric charge $Q$ is
\begin{equation}
	\label{eq:hypercharge_normalization}
	Q = T_3 + \frac{Y}{2}
\end{equation}
where $T_3$ is the third component of the weak isospin.
It should be noted that {\minlag} automatically treats neutral scalars (i.\,e.\ uncharged under \emph{all} – even global – charges) as real for simplicity.\footnote{%
	In the case of a scalar which is neutral with respect to all charges, a complex scalar is equivalent to two real scalars (and must be specified as such in the model file).%
}

In a bit more detail, the user model file must contain a list\footnote{%
	Generally, whenever a \q{list} is mentioned, any {\Python} iterable is allowed.
}
of \lstinline[language=Python]|BSMModel| objects.
These objects can be constructed using the \lstinline[language=Python]|BSMModel()| constructor which takes two mandatory and one optional arguments, \lstinline[language=Python]|BSMModel(name, fields, param_values)|:
\begin{description}
	\item[\textcode{name}] The first argument is simply a name for the model as a string.
	The name is arbitrary and used as an identifier for the model.
	This is the name that is used to refer to the model on the command line.
	\item[\textcode{fields}] The second argument is a list of fields and represents the model's field content.
	Scalar (spin~$0$) or fermion (spin $\sfrac{1}{2}$) fields are denoted using \lstinline[language=Python]|ScalarField()| or \lstinline[language=Python]|FermionField()|, respectively.
	It should be noted that all fermions are defined in terms of (two-component) left-handed Weyl spinors (see \cref{app:conventions}).
	Fields are defined in a similar manner to the model itself, with arguments \lstinline[language=Python]|Field(symbol, su2_multiplicity, Y, z2=-1, u1=[…])|:
	\begin{description}
		\item[\textcode{symbol}] A name for the field, which will appear in the resulting Lagrangian.
		\item[\textcode{su2\_multiplicity}] The value $n$ corresponding to the representation of $\SUIIL$ which the $n$-plet field is in (singlet, doublet or triplet).
		\item[\textcode{Y}] The $\UIY$ hypercharge.
		\item[\textcode{z2}] The global $\gZ_2$ \q{parity} $\pm 1$ for the field.
		If omitted, the default is \lstinline[language=Python]|z2=-1|.
		SM fields have \lstinline[language=Python]|z2=1|.
		\item[\textcode{u1} (optional)] A list of an arbitrary number $N$ of global $\gU(1)$ charges for the field.
		$N$ must be consistent across all fields.
		SM fields are assumed to be neutral under all global $\gU(1)$ charges.
	\end{description}
	\item[\textcode{param\_values} (optional)] The third argument can be used to define several models at once which have the same field content except for a constant shift in the hypercharge of all fields, as in \cite{Restrepo:2013aga}.
	In most cases, this argument can be omitted.
\end{description}
For example, the model T1-3-B, which is studied for $\alpha = 0$ in \cite{Fiaschi:2018rky}, is defined as
\begin{lstlisting}[language=Python, style=display-code]
BSMModel('T1-3-B', (
	FermionField("Ψ",  1, Y= 0),
	FermionField("ψ'", 2, Y= 1),
	ScalarField ("ϕ",  3, Y= 0),
	FermionField("ψ",  2, Y=-1),
), (0, 2)), # α = -2 is equivalent to α = 2
\end{lstlisting}

As mentioned before, {\minlag} will automatically make some fermions vector-like if necessary in order to cancel gauge anomalies when processing a model.
In this case, a warning will be emitted to the user.
If this automatic modification is not desired, the model should explicitly be defined (with additional fermion fields) in such a way that there are no gauge anomalies.

\subsection{Use as a {\Python} package}
Once {\minlag} has been installed as described in \cref{sec:install}, its component {\Python} modules are in principle available for import in other {\Python} scripts.
All modules are contained within the \lstinline|min_lag| package and can be imported as demonstrated in \path{minimal-lagrangians.py}, the main executable file.
In this way, the data structures used and returned by the code can be inspected and used directly and dynamically if desired.
Note, however, that there are currently no plans to provide a stable programming interface (e.\,g.\ class names and attributes, function or method names, or the number and names of their parameters) for use of the package in this manner.

\section{Program output and examples}
\label{sec:output}
The program only prints the potential involving at least one new (i.\,e.\ non-SM) field, that is, the Standard Model Lagrangian $\mathcal{L}_{\mtext{SM}}$ and the kinetic terms $\mathcal{L}_{\mtext{kin}}$ (which always have the same form) are omitted.
Explicitly, given a Lagrangian $\mathcal{L}$ of the form
\begin{equation}
	\mathcal{L} = \mathcal{L}_{\mtext{SM}} + \mathcal{L}_{\mtext{kin}} + V_{\mtext{BSM}}
\end{equation}
{\minlag} will output the most general renormalizable BSM potential $V_{\mtext{BSM}}$.
The mathematical notation in the program's output is explained in \cref{app:conventions,app:spinors}.
As indicated in the program's help message, the Lagrangian can be output in three different formats, which are described in the following.

Two models will be illustrated as examples in the following subsections: the model T1-1-A with $\alpha = 0$ from \cite{Restrepo:2013aga} (also studied in \cite{Farzan:2009ji}), and the singlet–triplet scalar model (\q{model~C}) from \cite{Cheung:2013dua}.%
\footnote{%
	As mentioned before, {\minlag} makes use of a number of identities (\cref{app:identities}) in order to eliminate redundant parameters from the Lagrangian.
	Since it has to make a particular and consistent choice about which terms to eliminate, some of the examples shown in this section do not exactly match the form given in the corresponding references.
	However, it can easily be verified that the generated Lagrangians are completely equivalent.%
}
They are defined as follows:
\begin{lstlisting}[language=Python, style=display-code]
	# arXiv:1308.3655 model T1-1-A
	BSMModel('T1-1-A', (
		ScalarField ("φ",  1, Y= 0),
		ScalarField ("ϕ'", 2, Y=-1),
		FermionField("ψ",  1, Y= 0),
		ScalarField ("ϕ",  2, Y= 1),
	), (0, 2)), # α = -2 is equivalent to α = 2
	# arXiv:1311.5896 model C: singlet-triplet scalar
	BSMModel('STS', (
		ScalarField ('S',  1, Y= 0),
		ScalarField ('T',  3, Y= 0),
	)),
\end{lstlisting}

\subsection{Plain text}
\label{sec:output-plain}
By default, the program will output the Lagrangian in plain text to the command-line terminal for a clearer and more compact presentation which does not require a {\LaTeX} processor.
This output format makes heavy use of Unicode \cite{Unicode} for optimal readability.
For example, \lstinline[language=sh]|minimal-lagrangians --omit-equivalent-scalars T1-1-A 0| prints the Lagrangian for the model T1-1-A with $\alpha = 0$ in the following form:\footnote{%
	For $\alpha = 0$, the model T1-1-A can be defined using only three new multiplets (instead of four) because the scalar doublets are conjugates of each other \cite{Restrepo:2013aga}.
	The option \lstinline[language=sh]|--omit-equivalent-scalars| is used in the examples to use the minimal case of three new multiplets.
	This is also the version of the model studied in \cite{Farzan:2009ji}.%
}%
\begin{lstlisting}[style=plain-text, autogobble=false]
 - M_ϕ'² ϕ'^† ϕ' - ½ M_φ² φ²
 - (λ₁ (H ϕ') φ + H.c.)
 - λ₂ (H^† H) (ϕ'^† ϕ') - λ₃ (H^† ϕ') (ϕ'^† H) - λ₄ (ϕ'^† ϕ')² - λ₅ (H^† H) φ² - λ₆ (ϕ'^† ϕ') φ² - (λ₇ (H ϕ')² + H.c.) - λ₈ φ⁴
 - (½ M_ψ ψ ψ + H.c.)
 - (y₁ (ϕ'^† L) ψ + H.c.)
\end{lstlisting}

As another example, running the command \lstinline[language=sh]|minimal-lagrangians STS| prints the Lagrangian for the singlet–triplet scalar model:
\begin{lstlisting}[style=plain-text, autogobble=false]
 - ½ M_T² Tr(T²) - ½ M_S² S²
 - λ₁ H^† T² H - λ₂ (H^† T H) S - λ₃ (H^† H) S² - λ₄ Tr(T²)² - λ₅ Tr(T²) S² - λ₆ S⁴
\end{lstlisting}

\subsection{\LaTeX}
\label{sec:output-latex}
Simple text is not the only implemented output format.
Instead of (Unicode) plain text, {\minlag} can also output {\LaTeX} commands to typeset the generated Lagrangians.
Using the same examples as before, \lstinline[language=sh]|minimal-lagrangians --omit-equivalent-scalars --format LaTeX T1-1-A 0| results in
\begin{align}
	&- M_{\phi'}^{2} {\phi'}^\dagger \phi' - \frac{1}{2} M_{\varphi}^{2} {\varphi}^2
	\notag
	\displaybreak[0]
	\\
	&- \left(\lambda_{1} \left(H \phi'\right) \varphi + \text{H.\,c.}\right)
	- \lambda_{2} \left({H}^\dagger H\right) \left({\phi'}^\dagger \phi'\right)
	- \lambda_{3} \left({H}^\dagger \phi'\right) \left({\phi'}^\dagger H\right)
	\notag
	\displaybreak[0]
	\\
	&\qquad
	- \lambda_{4} \left({\phi'}^\dagger \phi'\right)^2
	- \lambda_{5} \left({H}^\dagger H\right) {\varphi}^2 - \lambda_{6} \left({\phi'}^\dagger \phi'\right) {\varphi}^2
	- \left(\lambda_{7} \left(H \phi'\right)^2 + \text{H.\,c.}\right)
	\notag
	\displaybreak[0]
	\\
	&\qquad- \lambda_{8} {\varphi}^4
	\notag
	\displaybreak[0]
	\\
	&- \left(\frac{1}{2} M_{\psi} \psi \psi + \text{H.\,c.}\right)
	- \left(y_{1} \left({\phi'}^\dagger L\right) \psi + \text{H.\,c.}\right)
\end{align}
and \lstinline[language=sh]|minimal-lagrangians --format LaTeX STS| yields
\begin{align}
	&- \frac{1}{2} M_{T}^{2} \operatorname{Tr}\left({T}^2\right) - \frac{1}{2} M_{S}^{2} {S}^2
	\notag
	\displaybreak[0]
	\\
	&- \lambda_{1} {H}^\dagger {T}^2 H - \lambda_{2} \left({H}^\dagger T H\right) S - \lambda_{3} \left({H}^\dagger H\right) {S}^2 - \lambda_{4} \operatorname{Tr}\left({T}^2\right)^2
	\notag
	\displaybreak[0]
	\\
	&\qquad
	- \lambda_{5} \operatorname{Tr}\left({T}^2\right) {S}^2 - \lambda_{6} {S}^4
\end{align}
(some line breaks added manually).\footnote{%
	One method to automatically add line breaks to long equations is provided by the {\LaTeX} package \textcode{breqn}.%
}

\subsection{{\SARAH} model files}
\label{sec:output-sarah}
Finally, the third output format supported by {\minlag} allows one to generate model files for the tool {\SARAH} \cite{Staub:2013tta,Staub:2015kfa}, which can then be used to study the model in detail and subsequently generate code and model files for a large number of particle physics tools, such as {\SPheno} \cite{Porod:2003um,Porod:2011nf} and {\micrOMEGAs} \cite{Belanger:2001fz,Belanger:2018mqt}.
In {\SARAH}, one of the main tasks in implementing a model is specifying the Lagrangian, along with explicitly defining the components of all fields and every used parameter, all of which has to be done manually.
{\minlag} eliminates these steps by generating the most general renormalizable Lagrangian automatically and creating all the files needed by {\SARAH}.
Together, these programs form a tool chain which, after specifying a model's field content, largely automates the programmatic implementation of the model's details and rapidly yields executable code to calculate physical observables (see \cref{sec:tool_chain}).
An example making use of this tool chain is the analysis of the model T1-3-B ($\alpha = 0$) performed in \cite{Fiaschi:2018rky}.

In particular, the generated {\SARAH} model files also contain all information about vacuum expectation values (VEVs) potentially acquired by scalar fields and mixing of fields to new mass eigenstates after electroweak symmetry breaking (EWSB).
While EWSB does not affect the Lagrangian itself, the treatment of mixing is necessary when {\SARAH} is used to compute physical observables using {\SARAH} or its supported tools (like {\SPheno} and {\micrOMEGAs}).

For the example of the model T1-1-A ($\alpha = 0$) used in the previous \cref{sec:output-plain,sec:output-latex}, the relevant portions of the {\SARAH} model files generated by {\minlag}, which can be created using
\begin{lstlisting}[language=sh, style=display-code, numbers=none]
	minimal-lagrangians --omit-equivalent-scalars --sarah-no-scalar-cpv --format SARAH  T1-1-A 0
\end{lstlisting}
are shown in \cref{app:sarah-output}.\footnote{%
	The option \lstinline[language=sh]|--sarah-no-scalar-cpv| is included to enforce separate mixing matrices for neutral scalar and pseudoscalar components.
	Only defining a single mixing for all neutral scalars can cause issues in \SARAH's output routines.%
}
Unfortunately, depending on the names chosen for the fields, some characters must be replaced by {\minlag} (e.\,g.\ \lstinline|psi| in place of $\psi$) to conform to \SARAH's syntax rules.

In the main model file, all fields used in the input for {\minlag} are defined as required by \SARAH, including their $\SUIIL$ components:
\begin{lstlisting}[language=Mathematica, style=display-code]
FermionFields[[6]] = {psi, 1, psi0, 0, 1, 1, -1};
ScalarFields[[2]]  = {varphi, 1, varphi0, 0, 1, 1, -1};
ScalarFields[[3]]  = {phiq, 1, {phiq0, phiqm}, -1/2, 2, 1, -1};
RealScalars = {varphi};
\end{lstlisting}
In addition to the Lagrangian, information about the mixing of fields and their mixing matrices must now also be given in this main model file, and is written by {\minlag} as follows:
\begin{lstlisting}[language=Mathematica, style=display-code]
DEFINITION[EWSB][MatterSector] = {
	(* Standard Model mixing *)
	{{{uL}, {conj[uR]}}, {{UL, Vu}, {UR, Uu}}},
	{{{dL}, {conj[dR]}}, {{DL, Vd}, {DR, Ud}}},
	{{{eL}, {conj[eR]}}, {{EL, Ve}, {ER, Ue}}},
	{{vL}, {VL, Uneu}},
	(* mixing of new fields *)
	{{varphi0, phiq0Re}, {smx0, ZZs1}}
};
\end{lstlisting}
Any mixing involving the quarks, i.\,e.\ the Cabibbo–Kobayashi–Maskawa (CKM) matrix, is unaffected by the new fields in {\minlag} (which must be color-neutral) and is thus identical to the Standard Model in the {\SARAH} model files (as indicated using comment syntax).
In the case of this example (model T1-1-A with $\alpha = 0$), there are no new fields which mix with the neutrinos, so their mixing is also unaffected at tree level.
However, the new neutral scalar singlet $\varphi$ and the (real part of the) neutral component of the scalar doublet $\phi'$ both have the same quantum numbers after EWSB and thus mix to a new neutral scalar state with a mixing matrix (called \lstinline|smx0| and \lstinline|ZZs1|, respectively, in the generated {\SARAH} model file).
Note once again that all these details are determined automatically by {\minlag} using the simple definition
\begin{lstlisting}[language=Python, style=display-code]
	BSMModel('T1-1-A', (
		ScalarField ("φ",  1, Y= 0),
		ScalarField ("ϕ'", 2, Y=-1),
		FermionField("ψ",  1, Y= 0),
		ScalarField ("ϕ",  2, Y= 1),
	), (0, 2)),
\end{lstlisting}
and do not need to be specified by the user.

\section{Summary of the implementation}
The way in which fields, terms and the Lagrangian as a whole are represented internally in {\minlag} is very straightforward and simplistic.
Fields are objects whose properties are their mathematical symbol and their quantum numbers:
\begin{itemize}
	\item \lstinline|type|: scalar or fermion;
	\item \lstinline|su2_multiplicity|: the dimension of their representation under the gauge group $\SUIIL$ – the values $1$, $2$ and $3$ (singlets, doublets and triplets) are supported;
	\item \lstinline|hypercharge|: the charge under the gauge group $\UIY$;
	\item \lstinline|z2|: the parity ($\pm 1$) under the global $\gZ_2$ symmetry;
	\item (optional) \lstinline|u1|: the list of charges under the $N$ global $\gU(1)$ symmetries.
\end{itemize}
Terms are then essentially lists of such field objects, with code in place to ensure a consistent order and grouping of fields within a term.
A Lagrangian is then an (ordered) set of such terms.

An alternative to this very basic approach could have been to adapt a symbolic computation package with some additional rules for equivalence of terms, ordering and output formatting.
However, for the Lagrangian of a minimal model, the only required operations are multiplication of fields and addition of terms, where the latter is even not really needed because none of the terms of the final Lagrangian can be simplified by addition.
Consequently, a symbolic computation package would not have made the implementation much easier beyond providing the commutative property $\phi_1 \phi_2 = \phi_2 \phi_1$ in a product.
On the other hand, adapting an existing library for a purpose such as this, which it was not designed for, would take a significant amount of effort.

In a sense, {\minlag} works at a higher level of abstraction – it does not think in terms of individual variables in a product, but really only cares about the terms as a whole.
No operations are performed on these terms and the desired concept of \q{equality} of terms is not a precise one down to each constant factor, because these just affect the arbitrary definition of each term's coupling parameter.
Such details are of no interest for the program's purpose, rather complicating the decision of whether two terms should be treated as equal.
Only if integration with other symbolic computation tools is desired would it be useful to investigate whether employing such a package as a lower-level component is worthwhile.

The main component implementing the generation of all possible gauge- and Lorentz-invariant terms
is the method \lstinline[language=Python]|BSMModel.lagrangian|, along with the methods \lstinline[language=Python]|is_valid|, \lstinline[language=Python]|generate_terms| and \lstinline[language=Python]|filter_terms_identities| of the \lstinline[language=Python]|Model| class, which it uses to construct the Lagrangian.
The method \lstinline|BSMModel.lagrangian| works as follows:
First, a list containing all of the model's fields and their adjoints is created.
Then, all the possibilities of combining $n$ of these fields ($2 \le n \le 4$) are enumerated.
For example, given two real fields $\phi_1$ and $\phi_2$, this list would contain the combinations
$(\phi_1\phi_1\phi_1, \phi_1\phi_1\phi_2, \phi_1\phi_2\phi_2, \phi_2\phi_2\phi_2)$
for $n = 3$.
Note that the order of the fields does not matter.
However, only those combinations which can be used to form invariant terms at all are kept as candidates.

The check whether such a combination of fields can yield invariant terms is done by \lstinline[language=Python]|Model.is_valid|.
Making use of formalisms for Weyl spinors and $\gSU(2)$ multiplets (see \cref{app:spinors}), it is simple to determine whether this is the case:
Every lowered index must appear in a sum with a raised index of the same kind.
Since fermion fields have mass dimension $\massdim^{\sfrac{3}{2}}$, there can be at most two of them in a renormalizable term in any case, so this reduces to a check that they are both either left-handed or conjugate left-handed (i.\,e.\ right-handed) spinors.
In general, though, the number of indices must be even for each kind of index.
For $\gSU(2)$, an $n$-plet has $n - 1$ indices, so the sum
$\sum_i (n_i - 1)$ must be an even number.
For the abelian groups ($\gU(1)$, $\gZ_2$), the sum of each kind of charge must be zero: $\sum_i q_i = 0$.
Alternatively, since the value of the \q{parity group} $\gZ_2$ is usually given as $p_i = e^{iq_i} = \pm 1$ (\q{even/odd}), the product of all $\gZ_2$ values must be one: $\prod_i p_i = 1$.

\begin{table}
	\centering
	\caption{%
		Types of terms which are allowed in the interaction potential of a renormalizable theory for a set of bosons $\{\phi_i\}$ and fermions $\{\psi_i\}$.%
	}
	\label{tab:renormalizable_terms}
	\begin{tabular}{lccl}
		\toprule
		Term & Type $(N_{\mtext{s}}, N_{\mtext{f}})$ & Mass dimension & Description
		\\
		\midrule
		$\phi_i \phi_j$ & $(2, 0)$ & $\massdim^2$ & bosonic mass term
		\\
		$\psi_i \psi_j$ & $(0, 2)$ & $\massdim^3$ & fermionic mass term
		\\
		$\phi_i \phi_j \phi_k$ & $(3, 0)$ & $\massdim^3$ & cubic interaction
		\\
		$\phi_i \phi_j \phi_k \phi_m$ & $(4, 0)$ & $\massdim^4$ & quartic interaction
		\\
		$\phi_i \psi_j \psi_k$ & $(2, 1)$ & $\massdim^4$ & Yukawa interaction
		\\
		\bottomrule
	\end{tabular}
\end{table}

All the combinations of fields which are identified as potentially valid are then given to \lstinline[language=Python]|Model.generate_terms|.
This method contains the algorithm to determine, given an arbitrary collection of fields, what terms involving this specific combination must be added to the Lagrangian.
As mentioned before, Lorentz invariance is easy to obtain since the combination of fields will either contain no or two Weyl spinors, and in both cases there is only one way to match up the spinor indices.
For gauge and global invariance, $\SUIIL$, as the only non-abelian group, is the only one which could couple the fields in non-trivial ways.
Invariance under the abelian groups is automatically ensured because the charges were checked in the previous step.
In any case, a term is either not or automatically invariant under abelian groups – there is only one way to couple the fields.
Finally, \lstinline[language=Python]|Model.filter_terms_identities| determines if a term can be omitted due to the identities in \cref{app:identities}.

To recap, only three different representations of $\gSU(2)$ are used in the considered models (singlets, doublets and triplets).
Furthermore, the restriction to renormalizable Lagrangians limits the Lorentz structure of terms to the possibilities listed in \cref{tab:renormalizable_terms}.
Together, these assumptions ensure that there is only a finite number of different types of terms, so that the easiest way to enumerate all the possible terms for a given set of fields is to simply go through all the different cases.

\section{Numerical analysis tool chain}
\label{sec:tool_chain}
As mentioned before, {\minlag} adds another piece to a tool chain formed from existing particle physics code, allowing one to automate most of the model implementation starting just with a model's field content.

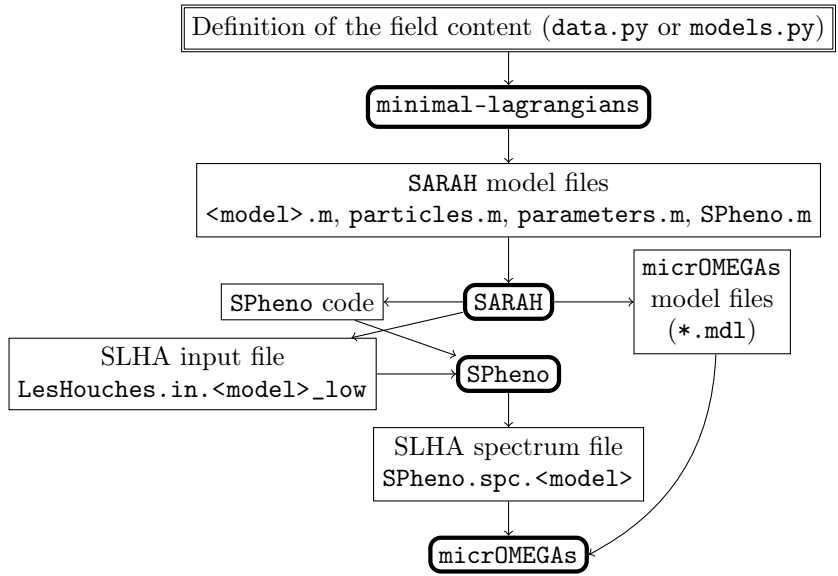
\begin{figure}
	\centering
	\input{res/flowchart.tex}
	\caption{%
		Flowchart illustrating the procedure to implement a model and run the numerical code within the computational tool chain started by \minlag.
		Boxes with rounded corners and thick borders represent programs, while the others represent files.
		The box with double-line borders indicates files which require user input.%
	}
	\label{fig:tool_chain}
\end{figure}

An illustration of how a model is implemented and analyzed using these tools is shown in \cref{fig:tool_chain}.
At the beginning, the model's field content is defined in \path{data.py} (or the user-supplied model file), as explained in \cref{sec:model_definition}.
{\minlag} can then be used to generate model files for {\SARAH} containing the most general renormalizable Lagrangian, as well as the definitions for potential VEVs or mixing of the fields after electroweak symmetry breaking.
Since {\SARAH} has a great number of features, it can sometimes be necessary or desirable to make some changes or additions to the generated model files.
With this potentially modified set of model files, {\SARAH} can be used to generate both the {\SPheno} code and model files for {\micrOMEGAs}.

At this point, the implementation of the model is already complete.
The generated code can be compiled and used to perform numerical calculations.
{\SPheno} takes a file in the SUSY Les Houches Accord (SLHA) format \cite{Skands:2003cj,Allanach:2008qq} as input.
In this file, the input parameters are provided and a number of settings (for example, which file formats are output, whether some calculations should be disabled and what conventions are used) can be customized.
{\SARAH} provides a template for this input file.
Running {\SPheno} then produces a spectrum file, also in SLHA format.
It contains the mass spectrum (including the mixing matrices) for the specified parameter point as well as lepton and quark flavor violation observables, some observables like $g - 2$ and (if enabled) branching ratios for particle decays.
In turn, this spectrum file can be used as an input to {\micrOMEGAs}, which extracts the mass spectrum and uses it to perform the calculation of dark matter observables.

\section{Verification and test suite}
{\minlag} contains a comprehensive test suite, which can be run using the \path{test.py} file contained within the program package.
Among other checks, this tests whether the program produces the correct Lagrangian for the following models, which have been compared with the literature:
\begin{itemize}
	\item T1-3-B with $\alpha = 0$, which is studied in \cite{Fiaschi:2018rky}.
	\item T1-1-A with $\alpha = 0$, as given in \cite{Farzan:2009ji} (which presents an implementation of this model).
	\item The simplified dark matter models given in \cite{Cheung:2013dua}:
	\begin{itemize}
		\item the singlet–doublet fermion model (SDF, \q{model~A});
		\item the singlet–doublet scalar model (SDS, \q{model~B});
		\item the singlet–triplet scalar model (STS, \q{model~C}).
	\end{itemize}
	\item The seesaw mechanism type~II, also called the Higgs triplet model (see e.\,g.\ \cite{Kanemura:2012rs}).
\end{itemize}
Additionally, it has been verified manually that the generated output is correct for a variety of different models, in particular for the seesaw mechanisms of types~I and III, and the model T1-2-A ($\alpha = 0$) studied in \cite{Esch:2018ccs}.

\section{Outlook}
{\minlag} will hopefully prove to be a useful tool both in accelerating phenomenological model building studies and providing automated, computer-assisted verification.
There are currently some restrictions to specific cases which could be generalized in the future.
The main limitation is the size of the $\gSU(2)$ representation, which cannot be larger than triplets.
The generalization to arbitrary representations would require a change in approach from a static list of conditions.
Moreover, the code currently only handles the Standard Model gauge group, so it is not possible to use models with new gauge fields.
The ultimate generalization would then be to not only handle representations of $\gSU(2)$, but of arbitrary $\gSU(N)$ gauge groups.
This would additionally require a general condition for the cancellation of gauge anomalies.

\section*{Acknowledgments}
This program resulted from work as part of the author's master's thesis at the Institute of Theoretical Physics at WWU Münster under the supervision of Michael Klasen.
My thanks extend to Michael Klasen for his guidance and valuable input on the manuscript as well as program features, and Karol Kovařík and Sonja Esch for many enlightening discussions during that time.

\clearpage

\appendix

\section{Conventions and notation}
\label{app:conventions}
In this work, all physical quantities are given in natural units, i.\,e.\ velocities are expressed in units of the speed of light in vacuum $c$ and actions or angular momenta are given in units of the reduced Planck constant $\hbar$.
The notation employed is then
\begin{equation}
	c = 1
	\qquad
	\hbar = 1
\end{equation}
with the physical dimension implicit in the seemingly dimensionless expressions on the right-hand side.

The Einstein summation convention is employed unless noted otherwise, with a sum over any two matching indices.
The position of the index does not, a priori, have any numerical meaning.

The normalization for the hypercharge $Y$, i.\,e.\ the charge corresponding to the $\UIY$ subgroup of the Standard Model gauge group $\SUIIIC \times \SUIIL \times \UIY$, was chosen such that
\begin{equation}
	Q = T_3 + \frac{Y}{2}
\end{equation}
(cf.\ \cref{eq:hypercharge_normalization}) where $Q$ is the electric charge and $T_3$ is an eigenvalue of the third $\SUIIL$ generator (\q{weak isospin}).

All fermion fields are left-handed Weyl spinors, i.\,e.\ $(\sfrac{1}{2}, 0)$ representations of $\gSL(2, \fieldC)$ (the covering group of the Lorentz group $\gSO(1, 3)^+$).
A four-component Dirac spinor can be constructed from a pair of two of such two-component spinors.
For illustration, this is the definition of the Standard Model's field content in \minlag:\footnote{%
	Technically, this only includes the first generation of the SM matter fields, since {\minlag} currently has no real concept of generation indices.
	However, the SM fermions are either singlets or doublets and can only appear in BSM Yukawa terms.
	There are no redundant term identities which are affected by different numbers of generations in this case, so one can simply add generation indices to \minlag's \q{one-generation} output in these cases.
	Again, the SM fields are assumed to be neutral under any global $\gU(1)$ charges.%
}
\begin{lstlisting}[language=Python, style=display-code]
STANDARD_MODEL = Model('Standard Model', (
	ScalarField ('H',     2, Y= 1,               z2=1),
	FermionField('L',     2, Y=-1,               z2=1),
	FermionField('Q',     2, Y=Fraction(' 1/3'), z2=1),
	FermionField('e_R^c', 1, Y= 2,               z2=1),
	FermionField('u_R^c', 1, Y=Fraction('-4/3'), z2=1),
	FermionField('d_R^c', 1, Y=Fraction(' 2/3'), z2=1),
))
\end{lstlisting}
Note that this does not include the SM gauge fields.
The fermion fields, split into their $\SUIIL$ representation components,\footnote{%
	As demonstrated in \cref{sec:output-sarah}, the handling of the $\SUIIL$ multiplet components shown in \cref{eq:sm-su2-components} is completely automated and does not require any user input.
	For example, by specifying \lstinline|FermionField('L', 2, Y=-1, z2=1)| as in the example for the Standard Model, {\minlag} determines that \lstinline|L| has two $\SUIIL$ components, whose electric charges are $0$ and $-1$, and will use this information to determine which fields mix after electroweak symmetry breaking.%
} are
\begin{equation}
	\label{eq:sm-su2-components}
	L = \begin{pmatrix}
		\nu_\upL \\ e_\upL
	\end{pmatrix},
	\quad
	Q = \begin{pmatrix}
		u_\upL \\ d_\upL
	\end{pmatrix},
	\quad
	e_\upR^\cconj,
	\quad
	u_\upR^\cconj,
	\quad
	d_\upR^\cconj
\end{equation}
where, once again, all fields are two-component $(\sfrac{1}{2}, 0)$ spinors.
The Standard Model Dirac fermions, i.\,e.\ the leptons and quarks, would thus take the form
\begin{equation}
	e = \begin{pmatrix}
		e_\upL \\ e_\upR = i\sigma_2 \ml(e_\upR^\cconj\mr)^*
	\end{pmatrix},
	\quad
	u = \begin{pmatrix}
		u_\upL \\ u_\upR = i\sigma_2 \ml(u_\upR^\cconj\mr)^*
	\end{pmatrix},
	\quad
	d = \begin{pmatrix}
		d_\upL \\ d_\upR = i\sigma_2 \ml(d_\upR^\cconj\mr)^*
	\end{pmatrix}
\end{equation}
with the second Pauli matrix $\sigma_2$.
The neutrinos $\nu$ remain massless Weyl fermions in the Standard Model.

\section{Spinor formalism}
\label{app:spinors}
For two-component Weyl spinors, the formalism described in \cite{Dreiner:2008tw} (van der Waerden notation) is assumed.
In particular, an $\gSL(2, \fieldC)$-invariant product of two left-handed Weyl spinors $\psi$ and $\chi$ is defined as
\begin{equation}
	\psi\chi
	= \psi^a \chi_a
	= \varepsilon^{ab} \psi_b \chi_a
\end{equation}
with the \q{spinor metric} $\varepsilon$, the two-dimensional Levi-Civita symbol
\begin{align}
	\label{eq:spinor_metric}
	\varepsilon &= (\varepsilon^{ab}) = i\sigma_2
	= \begin{pmatrix}
		0 & 1 \\
		-1 & 0
	\end{pmatrix}
	\\
	\varepsilon^{-1} &= (\varepsilon_{ab}) = -i\sigma_2
	= \begin{pmatrix}
		0 & -1 \\
		1 & 0
	\end{pmatrix}
\end{align}
and the second Pauli matrix $\sigma_2$.

A completely analogous formalism can be constructed for representations of $\gSU(2)$ instead of $\gSL(2, \fieldC)$.
The main difference is that only one \q{kind} of index is necessary, so there are no \q{dotted} indices.
This is because the Lie algebra $\gsl(2, \fieldC)$ of $\gSL(2, \fieldC)$ is simply the complexification of the Lie algebra $\gsu(2)$ of $\gSU(2)$.
Consequently, $\gsl(2, \fieldC)$, with its two inequivalent fundamental representations, essentially consists of two copies of $\gsu(2)$, which thus only requires a single kind of index.
The value of such a formalism is, as usual, that it becomes trivial to determine whether a given expression is invariant (in this case under $\gSU(2)$ gauge transformations).
This is the case when every index is paired with a counterpart.

Correspondingly, the $\gSU(2)$-invariant product of two $\gSU(2)$ doublets $D$ and $E$ is defined as
\begin{equation}
	DE
	= D^a E_a
	= \varepsilon^{ab} D_b E_a
	= (i\sigma_2 D) \cdot E
\end{equation}
using the same \q{spinor metric} as in \cref{eq:spinor_metric}, where the doublets \q{naturally} have lower indices.
$\cdot$ denotes the ordinary scalar product.
An important distinction has to be made when adjoints $D^\dagger$ of doublets are involved.
These already have the correct transformation behavior (upper indices) and do not require the application of $\varepsilon$ when combined with \q{ordinary} doublets:
\begin{align}
	D^\dagger E
	&= (D^\dagger)^a E_a
	\\
	D^\dagger E^\dagger
	&= (D^\dagger)^a (E^\dagger)_a
	= \varepsilon_{ab} (D^\dagger)^a (E^\dagger)^b
	= D^\dagger (-i\sigma_2 E^*)
\end{align}

It should be kept in mind that both the convention for products of $\gSU(2)$ multiplets and for products of two-component (Weyl) spinors are in effect simultaneously.
For instance, the SM Yukawa terms involving the up quarks would be written as
\begin{equation}
	H Q_i (u_\upR^\cconj)_j
	= \varepsilon^{ab} \varepsilon^{\alpha\beta} H_b (Q_i)_{a\beta} ((u_\upR^\cconj)_j)_\alpha
\end{equation}
with the Higgs doublet $H$, the quark doublets $Q_i$, the up-quark singlets $(u_\upR^\cconj)_i$ and where $a, b$ are $\gSU(2)$ component indices and $\alpha, \beta$ are (Lorentz) spinor indices.

$\gSU(2)$ triplets are analogous to Lorentz vectors.
They can be constructed from a tensor product of two doublets ($\surep{2} \otimes \surep{2} = \surep{3} \oplus \surep{1}$), $\gSU(2)$'s fundamental representation, similar to the construction of a Lorentz vector from a tensor product of two Weyl spinors ($(\sfrac{1}{2}, \sfrac{1}{2}) = (\sfrac{1}{2}, 0) \otimes (0, \sfrac{1}{2})$).
An $\gSU(2)$ triplet $\Delta$ can thus be viewed as a tensor with two indices (or a $2 \times 2$ matrix) for the purposes of defining an invariant product:
\begin{equation}
	\begin{pmatrix}
		\Delta_1 \\
		\Delta_2 \\
		\Delta_3
	\end{pmatrix}
	= \begin{pmatrix}
		\Delta^{\sfrac{Y}{2} + 1} \\
		\Delta^{\sfrac{Y}{2}} \\
		\Delta^{\sfrac{Y}{2} - 1}
	\end{pmatrix}
	\rightarrow
	\frac{1}{\sqrt{2}} \Delta^i \sigma_i
	= \begin{pmatrix}
		\frac{1}{\sqrt{2}} \Delta^{\sfrac{Y}{2}} & \Delta^{\sfrac{Y}{2} + 1} \\
		\Delta^{\sfrac{Y}{2} - 1} & -\frac{1}{\sqrt{2}} \Delta^{\sfrac{Y}{2}}
	\end{pmatrix}
\end{equation}
where the superscripts indicate the values of the component fields' electric charges given the hypercharge $Y$.

Since the triplets correspond to $\gSU(2)$'s adjoint representation, it is simple to determine the transformation behavior of the order-2 tensors.
Products between doublets and triplets should then be treated as ordinary matrix multiplications, with the requirement that a doublet multiplying from the left must have an upper index ($i\sigma_2 D$ or $D^\dagger$) and one multiplying from the right must have a lower index ($D$ or $-i\sigma_2 D^\dagger$).
To summarize, for two $\gSU(2)$ doublets $D$ and $E$ and a triplet $\Delta = \Delta^i \sigma_i$:
\begin{align}
	D \Delta E
	&= D^a \tensor{\Delta}{_a^b} E_b
	= (i\sigma_2 D) \cdot \Delta E
	\\
	D^\dagger \Delta E
	&= (D^\dagger)^a \tensor{\Delta}{_a^b} E_b
	= D^\dagger \Delta E
	\\
	D \Delta E^\dagger
	&= D^a \tensor{\Delta}{_a^b} (E^\dagger)_b
	= (i\sigma_2 D) \cdot \Delta (-i\sigma_2 E^*)
	\\
	D^\dagger \Delta E^\dagger
	&= (D^\dagger)^a \tensor{\Delta}{_a^b} E_b
	= D^\dagger \Delta (-i\sigma_2 E^*)
\end{align}
There is no distinction between $\Delta$ and $\Delta^\dagger$ concerning the defined products.

\section{Identities for redundant terms}
\label{app:identities}
In the process of finding the most general renormalizable Lagrangian, it becomes apparent that simply writing down all possible terms which are allowed by gauge and Lorentz invariance does not necessarily yield a minimal set of parameters.
If there exist identities linking different terms in the Lagrangian, it is possible to eliminate a term by expressing it through others via the identity and thus reduce the parameter space.
To be specific, given a Lagrangian
\begin{equation}
	\lagL = \sum_i \lambda_i A_i + R
\end{equation}
with parameters $\lambda_i$, products of fields $A_i$ and remaining terms $R$, if there is an identity of the form
\begin{equation}
	\sum_i A_i = 0
\end{equation}
then one of the terms $\lambda_i A_i$ can be omitted from the Lagrangian (equivalently, one can set $\lambda_i = 0$) without any loss of generality.

There is a number of such identities which is relevant to doublets and triplets of $\gSU(2)$ and thus the program's domain of operation.
The following is a list of these identities, which are taken into account by {\minlag} when constructing a Lagrangian.
Except for \cref{thm:doublets_triplet,thm:triplets_trilinear}, the discussion can be limited to scalar (or, in general, bosonic) fields because all quartic terms involving fermions are non-renormalizable.

\subsection{Identities}
Note that, as before, products between $\gSU(2)$ multiplets are defined as in \cref{app:spinors}.
The identities are generally still valid when replacing any of the fields by their corresponding adjoint fields.

\begin{identity}
	\label{thm:bosonic_doublet_product}
	For any bosonic $\gSU(2)$ doublet $D$:
	\begin{equation}
		D D
		= 0
	\end{equation}
\end{identity}

\begin{identity}
	\label{thm:doublets_quartic}
	For any two bosonic $\gSU(2)$ doublets $D_1$ and $D_2$:
	\begin{equation}
		|D_1|^2 |D_2|^2
		= |D_1 D_2|^2 + |D_1^\dagger D_2|^2
	\end{equation}
\end{identity}

\noindent
This implies that for $D_1 \ne D_2$, only \emph{two} of the three terms
\begin{enumerate}
	\item $|D_1|^2 |D_2|^2$
	\item $|D_1 D_2|^2$
	\item $|D_1^\dagger D_2|^2$
\end{enumerate}
are relevant to the parameter space.
For $D_1 = D_2 = D$, the second term is zero (\cref{thm:bosonic_doublet_product}), and thus $|D|^2 |D|^2 = |D^\dagger D|^2 = (D^\dagger D)^2$, so there is only \emph{one} relevant term in this case.

\begin{identity}
	\label{thm:doublets_triplet}
	For any two $\gSU(2)$ doublets $D_1$ and $D_2$ and any triplet $\Delta = \Delta^i \sigma_i$:
	\begin{equation}
		D_1 \Delta D_2
		= \begin{cases}
			D_2 \Delta D_1
			& \text{if at most one of the factors is fermionic}
			\\
			-D_2 \Delta D_1
			& \text{else}
		\end{cases}
	\end{equation}
\end{identity}

\begin{identity}
	\label{thm:triplets_trilinear}
	For any set of $\gSU(2)$ triplets $\Delta_i = \Delta_i^j \sigma_j$ ($i \in \{1, 2, 3\} = I$) and an arbitrary map of indices $\sigma: I \to I$,
	\begin{equation}
		\Tr(\Delta_{\sigma(i)} \Delta_{\sigma(j)} \Delta_{\sigma(k)})
		= \pm \varepsilon_\sigma \Tr(\Delta_i \Delta_j \Delta_k)
	\end{equation}
	where the negative sign can only occur in some cases if at least two of the triplets are fermionic and
	\begin{equation*}
		\varepsilon_\sigma
		= \begin{cases}
			1 &
			\text{if $\sigma$ is an even (\q{cyclic}) permutation}
			\\
			-1 &
			\text{if $\sigma$ is an odd (\q{anti-cyclic}) permutation}
			\\
			0 &
			\text{else}
		\end{cases}
	\end{equation*}
\end{identity}

\noindent
This means that, with another potential (arbitrary) factor $X$, only one of the terms $\Tr(\Delta_i \Delta_j \Delta_k) X$ with a certain permutation of $i, j, k$ can appear, and only if $i, j, k$ are pairwise different.

\begin{identity}
	\label{thm:triplets_quartic}
	For any set of $\gSU(2)$ triplets $\Delta_i = \Delta_i^j \sigma_j$ ($i \in \{1, 2, 3, 4\}$),
	\begin{align}
		&2\Tr\ml(\Delta_1 \Delta_2 \Delta_3 \Delta_4\mr)
		\notag\\
		&\quad=
		\Tr\ml(\Delta_1 \Delta_2\mr) \Tr\ml(\Delta_3 \Delta_4\mr)
		- \Tr\ml(\Delta_1 \Delta_3\mr) \Tr\ml(\Delta_2 \Delta_4\mr)
		+ \Tr\ml(\Delta_1 \Delta_4\mr) \Tr\ml(\Delta_2 \Delta_3\mr)
	\end{align}
\end{identity}

As a consequence of \cref{thm:triplets_quartic}, there is no need to include any traces of products of four triplets in the Lagrangian.
In other words, only terms of the form $\Tr(\Delta_i \Delta_j) \Tr(\Delta_k \Delta_m)$ are relevant to the parameter space.

To give a simple example, for a single bosonic $\gSU(2)$ triplet $\Delta$, only \emph{two} quartic terms must be included in a fully general Lagrangian, e.\,g.%
\begin{enumerate}
	\item $\Tr\ml(\Delta^\dagger \Delta\mr)^2$ and
	\item $\Tr\ml(\Delta^\dagger\mr)^2 \Tr\ml(\Delta\mr)^2$.
\end{enumerate}
If $\Delta$ is a real triplet ($\Delta^\dagger = \Delta$), it holds that $\Tr(\Delta^2)^2 = 2 \Tr(\Delta^4)$, so only \emph{one} term must be included in this case.

For any bosonic $\gSU(2)$ doublets $D_1$, $D_2$ and triplets $\Delta_1 = \Delta_1^i \sigma_i$, $\Delta_2 = \Delta_2^i \sigma_i$, the following identities hold:
\begin{identity}
	\label{thm:triplets_doublets_quartic_1}
	\begin{equation}
		D_1^\dagger \Delta_1 \Delta_2 D_2 + D_1^\dagger \Delta_2 \Delta_1 D_2
		= D_1^\dagger \acomm{\Delta_1}{\Delta_2} D_2
		= D_1^\dagger D_2 \Tr(\Delta_1 \Delta_2)
	\end{equation}
\end{identity}

\begin{identity}
	\label{thm:triplets_doublets_quartic_2}
	\begin{equation}
		D_1^\dagger \Delta_1 \Delta_2 D_2
		= -D_2 \Delta_2 \Delta_1 D_1^\dagger
	\end{equation}
\end{identity}

\Cref{thm:triplets_doublets_quartic_1} implies that, given appropriate hypercharges (otherwise, gauge invariance would be violated), only \emph{two} of the three terms
\begin{enumerate}
	\item $D_1^\dagger \Delta_1 \Delta_2 D_2 + \Hc$
	\item $D_1^\dagger \Delta_2 \Delta_1 D_2 + \Hc$
	\item $D_1^\dagger D_2 \Tr(\Delta_1 \Delta_2) + \Hc$
\end{enumerate}
are relevant to the parameter space.
If $\Delta_1 = \Delta_2 = \Delta$, it holds that $D_1^\dagger \Delta^2 D_2 = D_1^\dagger D_2 \Tr(\Delta^2)$, so there is only \emph{one} relevant term in this case.
Similarly, if $D_1^\dagger = D_2 = D$, then $D D \Tr(\Delta_1 \Delta_2) = 0$, leaving \emph{one} (non-zero) relevant term again.
\Cref{thm:triplets_doublets_quartic_2} ensures that any terms where the doublets are swapped can be omitted.

\subsection{Proofs}
The commutation and anti-commutation relations of the Pauli matrices
\begin{align}
	\label{eq:pauli_commutator}
	\comm{\sigma_i}{\sigma_j} &= 2i\varepsilon_{ijk} \sigma_k
	\\
	\label{eq:pauli_anticommutator}
	\acomm{\sigma_i}{\sigma_j}
	&= 2\delta_{ij} \idmat
	\\
	\label{eq:pauli_product}
	\Rightarrow \quad
	\sigma_i \sigma_j
	&= \delta_{ij} \idmat + i\varepsilon_{ijk} \sigma_k
	\intertext{as well as the following well-known trace identities}
	\label{eq:pauli_trace1}
	\Tr(\sigma_i) &= 0
	\\
	\label{eq:pauli_trace2}
	\Tr(\sigma_i \sigma_j)
	&= 2\delta_{ij}
	\\
	\label{eq:pauli_trace3}
	\Tr(\sigma_i \sigma_j \sigma_k)
	&= 2i\varepsilon_{ijk}
	\\
	\label{eq:pauli_trace4}
	\Tr(\sigma_i \sigma_j \sigma_k \sigma_m)
	&= 2 (\delta_{ij} \delta_{km} - \delta_{ik} \delta_{jm} + \delta_{im} \delta_{jk})
	\intertext{and the identity}
	\label{eq:identity_sigma2i2}
	\sigma_2 \sigma_i \sigma_2
	&= -\sigma_i^\transp
	= -\sigma_i^*
\end{align}
will be used here.

\begin{proof}[Proof of \cref{thm:bosonic_doublet_product}]
	\begin{equation*}
		DD
		= D^a D_a
		= \varepsilon^{ab} D_b D_a
		= -\varepsilon^{ba} D_a D_b
		= -D^b D_b
		= -DD
	\end{equation*}
\end{proof}

\begin{proof}[Proof of \cref{thm:doublets_quartic}]
	\begin{align*}
	|D_1|^2 |D_2|^2
	&= (|D_{11}|^2 + |D_{12}|^2) (|D_{21}|^2 + |D_{22}|^2)
	\\
	&= |D_{11}|^2 |D_{21}|^2 + |D_{12}|^2 |D_{21}|^2 + |D_{11}|^2 |D_{22}|^2 + |D_{12}|^2 |D_{22}|^2
	\displaybreak[0]
	\\
	|D_1^\transp i\sigma_2 D_2|^2
	&= |D_{12} D_{21} - D_{11} D_{22}|^2
	\\
	&= |D_{12} D_{21}|^2 + |D_{11} D_{22}|^2
	- D_{12}^\dagger D_{21}^\dagger D_{11} D_{22} - D_{11}^\dagger D_{22}^\dagger D_{12} D_{21}
	\displaybreak[0]
	\\
	|D_1^\dagger D_2|^2
	&= |D_{11}^\dagger D_{21} + D_{12}^\dagger D_{22}|^2
	\\
	&= |D_{11}|^2 |D_{21}|^2 + |D_{12}|^2 |D_{21}|^2
	+ D_{12}^\dagger D_{21}^\dagger D_{11} D_{22} + D_{11}^\dagger D_{22}^\dagger D_{12} D_{21}
	\end{align*}
\end{proof}

\begin{proof}[Proof of \cref{thm:doublets_triplet}]
	\begin{align*}
		D_1 \Delta D_2
		&= (D_1)^a \tensor{\Delta}{_a^b} (D_2)_b
		\\
		&= \varepsilon^{ac} (D_1)_c \Delta^i \tensor{(\sigma_i)}{_a^b} \varepsilon_{bd} (D_2)^d
		\\
		&= (D_1)_c \Delta^i (D_2)^d (-\varepsilon^{ca}  \tensor{(\sigma_i)}{_a^b} \varepsilon_{bd})
		\\
		&= (D_1)_c \Delta^i (D_2)^d \tensor{(-\varepsilon \sigma_i \varepsilon^{-1})}{^c_d}
		\\
		&= (D_1)_c \Delta^i (D_2)^d \tensor{(-\sigma_2 \sigma_i \sigma_2)}{^c_d}
		\\
		&\stackrel{\mathclap{\text{\eqref{eq:identity_sigma2i2}}}}{=}
		\;
		(D_1)_c \Delta^i (D_2)^d \tensor{(\sigma_i^\transp)}{^c_d}
		\intertext{The sign resulting from commuting $(D_2)^d$, $\Delta^i$ and $(D_1)_c$ depends on the number of fermions:}
		&= \pm (D_2)^d \Delta^i (D_1)_c  \tensor{(\sigma_i)}{_d^c}
		\\
		&= \pm (D_2)^d  \tensor{\Delta}{_d^c} (D_1)_c
		\\
		&= \pm D_2 \Delta D_1
	\end{align*}
\end{proof}

\begin{proof}[Proof of \cref{thm:triplets_trilinear}]
	\begin{align*}
		\Tr(\Delta_{\sigma(i)} \Delta_{\sigma(j)} \Delta_{\sigma(k)})
		&= \Delta_{\sigma(i)}^l \Delta_{\sigma(j)}^m \Delta_{\sigma(k)}^n \Tr(\sigma_l \sigma_m \sigma_n)
		\\
		&\stackrel{\mathclap{\text{\eqref{eq:pauli_trace3}}}}{=}
		\;
		2i \Delta_{\sigma(i)}^l \Delta_{\sigma(j)}^m \Delta_{\sigma(k)}^n \varepsilon_{lmn}
		\intertext{
			The indices $l, m, n$ can be renamed in such a way that this takes the form $\Delta_i^{l'} \Delta_j^{m'} \Delta_k^{n'} \varepsilon_{\sigma(l'm'n')}$, where the order of indices on $\varepsilon$ is determined by $\sigma$.
			If some of the triplets are fermionic, this can introduce an additional negative sign depending on which factors have to be commuted:
		}
		&= \pm 2i \Delta_i^{l'} \Delta_j^{m'} \Delta_k^{n'} \varepsilon_{\sigma(l'm'n')}
		\\
		&= \pm 2i \Delta_i^{l'} \Delta_j^{m'} \Delta_k^{n'} \varepsilon_\sigma \varepsilon_{l'm'n'}
		\\
		&= \pm \varepsilon_\sigma \Tr(\Delta_i \Delta_j \Delta_k)
	\end{align*}
\end{proof}

\begin{proof}[Proof of \cref{thm:triplets_quartic}]
	\begin{align*}
		\Tr\ml(\Delta_a \Delta_b\mr) \Tr\ml(\Delta_c \Delta_d\mr)
		&= \Tr\ml(\Delta_a^i \Delta_b^j \sigma_i \sigma_j\mr) \Tr\ml(\Delta_c^k \Delta_d^m \sigma_k \sigma_m\mr)
		\\
		&= \Delta_a^i \Delta_b^j \Delta_c^k \Delta_d^m \Tr\ml(\sigma_i \sigma_j\mr) \Tr\ml(\sigma_k \sigma_m\mr)
		\\
		&\stackrel{\mathclap{\text{\eqref{eq:pauli_trace2}}}}{=}
		\;
		4 \Delta_a^i \Delta_b^j \Delta_c^k \Delta_d^m \delta_{ij} \delta_{km}
		\\
		&= 4 \ml(\Delta_a^i {\Delta_b}_i\mr) \ml(\Delta_c^j {\Delta_d}_j\mr)
		\displaybreak[0]
		\\[\medskipamount]
		\Tr\ml(\Delta_1 \Delta_2 \Delta_3 \Delta_4\mr)
		&= \Tr\ml(\Delta_1^i \Delta_2^j \Delta_3^k \Delta_4^m \sigma_i \sigma_j \sigma_k \sigma_m\mr)
		\\
		&= \Delta_1^i \Delta_2^j \Delta_3^k \Delta_4^m \Tr\ml(\sigma_i \sigma_j \sigma_k \sigma_m\mr)
		\\
		&\stackrel{\mathclap{\text{\eqref{eq:pauli_trace4}}}}{=}
		\;
		2 \Delta_1^i \Delta_2^j \Delta_3^k \Delta_4^m (\delta_{ij} \delta_{km} - \delta_{ik} \delta_{jm} + \delta_{im} \delta_{jk})
		\\
		&= 2 \Big(\ml(
			\Delta_1^i {\Delta_2}_i\mr) \ml(\Delta_3^j {\Delta_4}_j\mr)
		\\
		&\qquad
			- \ml(\Delta_1^i {\Delta_3}_i\mr) \ml(\Delta_2^j {\Delta_4}_j\mr)
			+ \ml(\Delta_1^i {\Delta_4}_i\mr) \ml(\Delta_2^j {\Delta_3}_j\mr)\Big)
	\end{align*}
\end{proof}

\begin{proof}[Proof of \cref{thm:triplets_doublets_quartic_1}]
	\begin{align*}
		D_1^\dagger \acomm{\Delta_1}{\Delta_2} D_2
		&= D_1^\dagger \Delta_1^i \Delta_2^j \acomm{\sigma_i}{\sigma_j} D_2
		\\
		&\stackrel{\mathclap{\text{\eqref{eq:pauli_anticommutator}}}}{=}
		\;
		2 D_1^\dagger \Delta_1^i \Delta_2^j \delta_{ij} \idmat D_2
		\\
		&\stackrel{\mathclap{\text{\eqref{eq:pauli_trace2}}}}{=}
		\;
		D_1^\dagger \Delta_1^i \Delta_2^j \Tr(\sigma_i \sigma_j) D_2
		\\
		&= D_1^\dagger \Tr(\Delta_1 \Delta_2) D_2
		\\
		&= D_1^\dagger D_2 \Tr(\Delta_1 \Delta_2)
	\end{align*}
\end{proof}

	\begin{proof}[Proof of \cref{thm:triplets_doublets_quartic_2}]
	\begin{align*}
		D_1^\dagger \Delta_1 \Delta_2 D_2
		&= (D_1^\dagger)^a \tensor{(\Delta_1)}{_a^b} \tensor{(\Delta_2)}{_b^c} (D_2)_c
		\\
		&= \varepsilon^{am} (D_1^\dagger)_m \Delta_1^i \tensor{(\sigma_i)}{_a^b} \Delta_2^j \tensor{(\sigma_j)}{_b^c} \varepsilon_{cn} (D_2)^n
		\\
		&= (D_2)^n \Delta_2^j \Delta_1^i (D_1^\dagger)_m
		\varepsilon^{am} \tensor{(\sigma_i)}{_a^b} \tensor{(\sigma_j)}{_b^c} \varepsilon_{cn}
		\\
		&= -(D_2)^n \Delta_2^j \Delta_1^i (D_1^\dagger)_m
		\tensor{(\sigma_2 \sigma_i \sigma_j \sigma_2)}{^m_n}
	\end{align*}
	Considering the factor $\sigma_2 \sigma_i \sigma_j \sigma_2$:
	\begin{align*}
		\sigma_2 \sigma_i \sigma_j \sigma_2
		&\stackrel{\mathclap{\text{\eqref{eq:pauli_product}}}}{=}
		\;
		\sigma_2 (\delta_{ij} \idmat + \varepsilon_{ijk} \sigma_k) \sigma_2
		\\
		&= \delta_{ij} \sigma_2^2 + i\varepsilon_{ijk} \sigma_2 \sigma_k \sigma_2
		\\
		&\stackrel{\mathclap{\text{\eqref{eq:identity_sigma2i2}}}}{=}
		\;
		\delta_{ij} \idmat - i\varepsilon_{ijk} \sigma_k^*
		\\
		&= (\sigma_i \sigma_j)^*
		\\
		&= (\sigma_j \sigma_i)^\transp
	\end{align*}
	Inserting this into the expression for $D^\dagger \Delta_1 \Delta_2 D$:
	\begin{align*}
		D_1^\dagger \Delta_1 \Delta_2 D_2
		&= -(D_2)^n \Delta_2^j \Delta_1^i (D_1^\dagger)_m
		\tensor{\ml((\sigma_j \sigma_i)^\transp\mr)}{^m_n}
		\\
		&= -(D_2)^n \Delta_2^j \Delta_1^i (D_1^\dagger)_m
		\tensor{(\sigma_j \sigma_i)}{_n^m}
		\\
		&= -(D_2)^n \Delta_2^j \tensor{(\sigma_j)}{_n^a} \Delta_1^i \tensor{(\sigma_i)}{_a^m} (D_1^\dagger)_m
		\\
		&= -(D_2)^n \tensor{(\Delta_2)}{_n^a} \tensor{(\Delta_1)}{_a^m} (D_1^\dagger)_m
		\\
		&= -D_2 \Delta_2 \Delta_1 D_1^\dagger
	\end{align*}
\end{proof}

\section{Example {\SARAH} output}
\label{app:sarah-output}
\begin{lstlisting}[language=Mathematica, style=display-code, title={\textcode{T1\_1\_A\_alpha\_0.m}}]
(* This file has been automatically generated by minimal-lagrangians *)
Off[General::spell];

Model`Name = "T1_1_A_alpha_0";
Model`NameLaTeX = "T1-1-A (\\alpha = 0)";
Model`Authors = "minimal-lagrangians (automatically generated)";
Model`Date = "2020-10-06";

(*-------------------------------------------*)
(*   particle content                        *)
(*-------------------------------------------*)

(* global symmetries *)
(* discrete ℤ₂ symmetry *)
Global[[1]] = {Z[2], Z2};

(* gauge groups *)
Gauge[[1]] = {B,   U[1], hypercharge, g1, False, 1};
Gauge[[2]] = {WB, SU[2], left,        g2, True,  1};
Gauge[[3]] = {G,  SU[3], color,       g3, False, 1};

(* matter fields *)
(*                   {name, gens, components, Y/2,  SU(2), SU(3), global} *)
(* Standard Model *)
FermionFields[[1]] = {q,    3,    {uL, dL},    1/6, 2,      3,    1};
FermionFields[[2]] = {l,    3,    {vL, eL},   -1/2, 2,      1,    1};
FermionFields[[3]] = {u,    3,    conj[uR],   -2/3, 1,     -3,    1};
FermionFields[[4]] = {d,    3,    conj[dR],    1/3, 1,     -3,    1};
FermionFields[[5]] = {e,    3,    conj[eR],      1, 1,      1,    1};

ScalarFields[[1]]  = {H,    1,    {Hp, H0},    1/2, 2,      1,    1};

(* new fields *)
FermionFields[[6]] = {psi, 1, psi0, 0, 1, 1, -1};
ScalarFields[[2]]  = {varphi, 1, varphi0, 0, 1, 1, -1};
ScalarFields[[3]]  = {phiq, 1, {phiq0, phiqm}, -1/2, 2, 1, -1};
RealScalars = {varphi};

(*----------------------------------------------*)
(*   DEFINITION                                 *)
(*----------------------------------------------*)
NameOfStates = {GaugeES, EWSB};

(* ----- before EWSB ----- *)
DEFINITION[GaugeES][LagrangianInput] = {
	(* Standard Model Lagrangian *)
	{LagNoHC,    {AddHC -> False}},
	{LagHC,      {AddHC -> True }},
	(* Lagrangian involving the new fields *)
	{LagBSMNoHC, {AddHC -> False}},
	{LagBSMHC,   {AddHC -> True }}
};

(* Standard Model Lagrangian *)
LagNoHC    = mu2 conj[H].H - 1/2 λ conj[H].H.conj[H].H;
LagHC      = -Yu u.q.H - Yd conj[H].d.q - Ye conj[H].e.l;

(* Lagrangian involving the new fields *)
LagBSMNoHC = - Mphiq2 conj[phiq].phiq - 1/2 Mvarphi2 varphi.varphi \
	 - lambda2 conj[H].H.conj[phiq].phiq - lambda3 conj[H].phiq.conj[phiq].H - lambda4 conj[phiq].phiq.conj[phiq].phiq - lambda5 conj[H].H.varphi.varphi - lambda6 conj[phiq].phiq.varphi.varphi - lambda8 varphi.varphi.varphi.varphi;
LagBSMHC = - lambda1 H.phiq.varphi \
	 - lambda7 H.phiq.H.phiq \
	 - 1/2 Mpsi psi.psi \
	 - y1 conj[phiq].l.psi;

(* ----- after EWSB ----- *)
(* gauge sector mixing *)
DEFINITION[EWSB][GaugeSector] = {
	{{VB,     VWB[3]}, {VP, VZ}, ZZ},
	{{VWB[1], VWB[2]}, {VWp, conj[VWp]}, ZW}
};

(* VEVs *)
DEFINITION[EWSB][VEVs] = {
	(* Standard Model Higgs VEV *)
	{H0,
		{v, 1/Sqrt[2]},
		{Ah, \[ImaginaryI]/Sqrt[2]},
		{hh, 1/Sqrt[2]}
	},
	(* BSM VEVs and splitting neutral scalars into real and imaginary parts *)
	{phiq0,
		{0, 1/Sqrt[2]},
		{phiq0Im, \[ImaginaryI]/Sqrt[2]},
		{phiq0Re, 1/Sqrt[2]}
	}
};

(* mixing *)
DEFINITION[EWSB][MatterSector] = {
	(* Standard Model mixing *)
	{{{uL}, {conj[uR]}}, {{UL, Vu}, {UR, Uu}}},
	{{{dL}, {conj[dR]}}, {{DL, Vd}, {DR, Ud}}},
	{{{eL}, {conj[eR]}}, {{EL, Ve}, {ER, Ue}}},
	{{vL}, {VL, Uneu}},
	(* mixing of new fields *)
	{{varphi0, phiq0Re}, {smx0, ZZs1}}
};

(* Dirac spinors *)
DEFINITION[EWSB][DiracSpinors] = {
	(* Standard Model Dirac spinors *)
	Fu -> {UL, conj[UR]},
	Fd -> {DL, conj[DR]},
	Fe -> {EL, conj[ER]},
	Fv -> {VL, conj[VL]},
	(* new Dirac spinors *)
	Fpsi0 -> {psi0, conj[psi0]}
};
\end{lstlisting}

\vspace{4ex}

\begin{lstlisting}[language=Mathematica, style=display-code, title={\textcode{particles.m}}]
ParticleDefinitions[GaugeES] = {
	(* new fields *)
	{varphi0, {Description -> "BSM field φ⁰",
			OutputName -> "varphi0",
			ElectricCharge -> 0,
			LaTeX -> "{\\varphi^0}"
		}
	},
	{phiq0, {Description -> "BSM field ϕ'⁰",
			OutputName -> "phiq0",
			ElectricCharge -> 0,
			LaTeX -> "{\\phi'^0}"
		}
	},
	{phiqm, {Description -> "BSM field ϕ'⁻",
			OutputName -> "phiqm",
			ElectricCharge -> -1,
			LaTeX -> "{\\phi'^-}"
		}
	},
	{psi0, {Description -> "BSM field ψ⁰",
			OutputName -> "psi0",
			ElectricCharge -> 0,
			LaTeX -> "{\\psi^0}"
		}
	},
	...
};

ParticleDefinitions[EWSB] = {
	(* new fields *)
	{phiqm, {Description -> "BSM field ϕ'⁻ (EWSB)",
			OutputName -> "phiqm",
			PDG -> {900},
			FeynArtsNr -> 900,
			ElectricCharge -> -1,
			LaTeX -> "{\\phi'^-}"
		}
	},
	{phiq0Im, {Description -> "BSM field ϕ'⁰Im (EWSB)",
			OutputName -> "phiq0Im",
			PDG -> {901},
			FeynArtsNr -> 901,
			ElectricCharge -> 0,
			LaTeX -> "{\\phi'^0}Im"
		}
	},
	{smx0, {Description -> "BSM field smx⁰ (EWSB)",
			OutputName -> "smx0",
			PDG -> {902, 903},
			FeynArtsNr -> 902,
			ElectricCharge -> 0,
			LaTeX -> "{smx^0}"
		}
	},
	{Fpsi0, {Description -> "BSM field Fψ⁰ (EWSB)",
			OutputName -> "Fpsi0",
			PDG -> {904},
			FeynArtsNr -> 904,
			ElectricCharge -> 0,
			LaTeX -> "{F\\psi^0}"
		}
	},

	(* Neutrinos *)
	{Fv,   {Description -> "Neutrinos",
			Mass -> LesHouches
		}
	},
	...
};
\end{lstlisting}

\vspace{4ex}

\begin{lstlisting}[language=Mathematica, style=display-code, title={\textcode{parameters.m}}]
ParameterDefinitions = {
	(* new parameters *)
	(* BSM parameters in the Lagrangian *)
	{Mphiq2, {Description -> "BSM parameter M_ϕ'²",
			OutputName -> "Mphiq2",
			Real -> True,
			LesHouches -> {"T11AALPHA0", 10},
			LaTeX -> "M_{\\phi'}^{2}"
		}
	},
	{Mvarphi2, {Description -> "BSM parameter M_φ²",
			OutputName -> "Mvarphi2",
			Real -> True,
			LesHouches -> {"T11AALPHA0", 11},
			LaTeX -> "M_{\\varphi}^{2}"
		}
	},
	{lambda2, {Description -> "BSM parameter λ₂",
			OutputName -> "lam2",
			Real -> True,
			LesHouches -> {"T11AALPHA0", 12},
			LaTeX -> "\\lambda_{2}"
		}
	},
	{lambda3, {Description -> "BSM parameter λ₃",
			OutputName -> "lam3",
			Real -> True,
			LesHouches -> {"T11AALPHA0", 13},
			LaTeX -> "\\lambda_{3}"
		}
	},
	{lambda4, {Description -> "BSM parameter λ₄",
			OutputName -> "lam4",
			Real -> True,
			LesHouches -> {"T11AALPHA0", 14},
			LaTeX -> "\\lambda_{4}"
		}
	},
	{lambda5, {Description -> "BSM parameter λ₅",
			OutputName -> "lam5",
			Real -> True,
			LesHouches -> {"T11AALPHA0", 15},
			LaTeX -> "\\lambda_{5}"
		}
	},
	{lambda6, {Description -> "BSM parameter λ₆",
			OutputName -> "lam6",
			Real -> True,
			LesHouches -> {"T11AALPHA0", 16},
			LaTeX -> "\\lambda_{6}"
		}
	},
	{lambda8, {Description -> "BSM parameter λ₈",
			OutputName -> "lam8",
			Real -> True,
			LesHouches -> {"T11AALPHA0", 17},
			LaTeX -> "\\lambda_{8}"
		}
	},
	{lambda1, {Description -> "BSM parameter λ₁",
			OutputName -> "lam1",
			LesHouches -> {"T11AALPHA0", 18},
			LaTeX -> "\\lambda_{1}"
		}
	},
	{lambda7, {Description -> "BSM parameter λ₇",
			OutputName -> "lam7",
			LesHouches -> {"T11AALPHA0", 19},
			LaTeX -> "\\lambda_{7}"
		}
	},
	{Mpsi, {Description -> "BSM parameter M_ψ",
			OutputName -> "Mpsi",
			LesHouches -> {"T11AALPHA0", 20},
			LaTeX -> "M_{\\psi}"
		}
	},
	{y1, {Description -> "BSM parameter y₁",
			OutputName -> "y1",
			LesHouches -> {"T11AALPHA0", 21},
			LaTeX -> "y_{1}"
		}
	},

	(* BSM mixing matrices *)
	{Uneu, {Description -> "BSM mixing matrix Uneu",
			OutputName -> "Uneu",
			LesHouches -> "UNEU",
			LaTeX -> "Uneu"
		}
	},
	{ZZs1, {Description -> "BSM mixing matrix ZZs1",
			OutputName -> "ZZs1",
			LesHouches -> "ZZS1",
			LaTeX -> "ZZs1"
		}
	},
	...
}
\end{lstlisting}

\vspace{4ex}

\begin{lstlisting}[language=Mathematica, style=display-code, title={\textcode{SPheno.m}}]
(* This file has been automatically generated by minimal-lagrangians *)

OnlyLowEnergySPheno = True;

MINPAR = {
	{1, lambdaInput},
	{10, Mphiq2Input},
	{11, Mvarphi2Input},
	{12, lambda2Input},
	{13, lambda3Input},
	{14, lambda4Input},
	{15, lambda5Input},
	{16, lambda6Input},
	{17, lambda8Input},
	{18, lambda1Input},
	{19, lambda7Input},
	{20, MpsiInput},
	{21, y1Input}
};

BoundaryLowScaleInput = {
	(* Standard Model *)
	{λ, lambdaInput},
	(* BSM *)
	{Mphiq2, Mphiq2Input},
	{Mvarphi2, Mvarphi2Input},
	{lambda2, lambda2Input},
	{lambda3, lambda3Input},
	{lambda4, lambda4Input},
	{lambda5, lambda5Input},
	{lambda6, lambda6Input},
	{lambda8, lambda8Input},
	{lambda1, lambda1Input},
	{lambda7, lambda7Input},
	{Mpsi, MpsiInput},
	{y1, y1Input}
};

(* NOTE: DEFINITION[MatchingConditions] and ParametersToSolveTadpoles should be
         adjusted manually if there are BSM fields which acquire a VEV *)
DEFINITION[MatchingConditions] = {
	{Yu, YuSM},
	{Yd, YdSM},
	{Ye, YeSM},
	{g1, g1SM},
	{g2, g2SM},
	{g3, g3SM},
	{v, vSM}
};

ParametersToSolveTadpoles = {mu2};

ListDecayParticles = {Fu, Fd, Fe, Hp, hh, Fpsi0, phiq0Im, phiqm, smx0};
ListDecayParticles3B = {{Fu, "Fu.f90"}, {Fd, "Fd.f90"}, {Fe, "Fe.f90"}};
\end{lstlisting}

\clearpage
\bibliographystyle{elsarticle-num}
\bibliography{bib}

\end{document}

%% file: settings.tex
\newcommand*{\q}{\enquote}
\newcommand*{\textcode}{\texttt}

\newcommand*{\ml}{\mleft}
\newcommand*{\mr}{\mright}
\newcommand*{\mtext}[1]{{\operatorfont #1}}
\newcommand*{\cconj}{{\mtext{c}}}
\newcommand*{\transp}{{\mathsf{T}}}
\newcommand*{\Hc}{\mtext{H.\,c.}}
\DeclareMathOperator{\Tr}{Tr}
\newcommand*{\idmat}{\mathbbm{1}}

\newcommand*{\git}{\textcode{git}}
\newcommand*{\Python}{\textcode{Python}}
\newcommand*{\pip}{\textcode{pip}}
\newcommand*{\minlag}{\textcode{mi\-ni\-mal-la\-gran\-gians}}
\newcommand*{\SARAH}{\textcode{SARAH}}
\newcommand*{\SPheno}{\textcode{SPheno}}
\newcommand*{\micrOMEGAs}{\textcode{micrOMEGAs}}

\newcommand*{\comm}[2]{\ml[#1, #2\mr]}
\newcommand*{\acomm}[2]{\ml\{#1, #2\mr\}}
\newcommand*{\lagL}{\mathcal{L}}
\newcommand*{\fieldC}{\mathbb{C}}
\newcommand*{\upL}{{\mtext{L}}}
\newcommand*{\upR}{{\mtext{R}}}
\newcommand*{\gZ}{\mathbb{Z}}
\newcommand*{\gU}{\operatorname{U}}
\newcommand*{\gSU}{\operatorname{SU}}
\newcommand*{\gSO}{\operatorname{SO}}
\newcommand*{\gSL}{\operatorname{SL}}
\newcommand*{\gsu}{\mathfrak{su}}
\newcommand*{\gsl}{\mathfrak{sl}}
\newcommand*{\SUIIIC}{\gSU(3)_{\mtext{C}}}
\newcommand*{\SUIIL}{\gSU(2)_{\mtext{L}}}
\newcommand*{\UIY}{\gU(1)_{\mtext{Y}}}
\newcommand*{\surep}[1]{\mathbf{#1}}
\newcommand*{\massdim}{\mathsf{M}}

\newtheorem{identity}{Identity}
\crefname{identity}{identity}{identities}
\crefname{appendix}{}{}

%% file: listings_styles.tex
\newcommand*{\postbreak}{%
	\mbox{\textcolor{red}{$\hookrightarrow$}\space}%
}

\colorlet{keywordcolor}{purple!60!black}
\colorlet{emphcolor}{teal!70!black}
\colorlet{commentcolor}{green!40!black}
\colorlet{texcsicolor}{blue}
\colorlet{texcsiicolor}{blue!50!black}
\colorlet{bracescolor}{orange!70!black}
\colorlet{dollarcolor}{red}
\definecolor{stringcolor}{RGB}{47, 0, 255}
\definecolor{codebgcolor}{RGB}{245, 245, 245}

\lstdefinestyle{plain-text}{
	frame=tb,
	numbers=none,
	columns=flexible,
	postbreak=\postbreak,
}
\lstdefinestyle{plain-text-80col}{
	style=plain-text,
	basicstyle=\ttfamily\scriptsize,
}
\lstdefinestyle{display-sh}{
	language=sh,
	numbers=left,
	basicstyle=\ttfamily\color{white},
	backgroundcolor=\color{black},
	frame=none,
	framexleftmargin=0.5em,
	xleftmargin=0.5em,
	postbreak=\postbreak,
}
\lstdefinestyle{display-code}{
	numbers=left,
	frame=none,
	backgroundcolor=\color{codebgcolor},
	framexleftmargin=0.5em,
	xleftmargin=0.5em,
	postbreak=\postbreak,
}

%% file: res/flowchart.tex
\providecommand*{\minlag}{\texttt{minimal-lagrangians}}
\tikzstyle{program}=[rectangle, draw, fill=white, align=center, rounded corners, ultra thick]
\tikzstyle{file}=[rectangle, draw, fill=white, align=center]
\footnotesize
\begin{tikzpicture}[node distance=3ex and 3em]
	\node[file, double] (datapy) {Definition of the field content (\texttt{data.py} or \texttt{models.py})};
	\node[program, below=of datapy] (minimallag) {\minlag};
	\node[file, below=of minimallag] (SARAHmdl) {{\SARAH} model files\\\path{<model>.m}, \path{particles.m}, \path{parameters.m}, \path{SPheno.m}};
	\node[program, below=of SARAHmdl, yshift=-1ex] (SARAH) {\SARAH};
	\node[file, left=of SARAH] (SPhenocode) {{\SPheno} code};
	\node[file, right=of SARAH] (micrOMEGAsmdl) {\micrOMEGAs\\model files\\(\path{*.mdl})};
	\node[program, below=of SARAH] (SPheno) {\SPheno};
	\node[file, left=of SPheno] (SPhenoin) {SLHA input file\\\path{LesHouches.in.<model>_low}};
	\node[file, below=of SPheno] (SPhenospc) {SLHA spectrum file\\\path{SPheno.spc.<model>}};
	\node[program, below=of SPhenospc] (micrOMEGAs) {\micrOMEGAs};
	\draw[->] (datapy) -- (minimallag);
	\draw[->] (minimallag) -- (SARAHmdl);
	\draw[->] (SARAHmdl) -- (SARAH);
	\draw[->] (SARAH) -- (SPhenocode);
	\draw[->] (SARAH) -- (micrOMEGAsmdl);
	\draw[->] (SARAH) -- (SPhenoin);
	\draw[->] (SPhenocode) -- (SPheno);
	\draw[->] (SPhenoin) -- (SPheno);
	\draw[->] (SPheno) -- (SPhenospc);
	\draw[->] (micrOMEGAsmdl) to[bend left] (micrOMEGAs.east);
	\draw[->] (SPhenospc) -- (micrOMEGAs);
\end{tikzpicture}